\documentclass[preprint,5p,times,twocolumn]{my-elsarticle}

\usepackage{amssymb}
\usepackage{amsmath}

\biboptions{compress}

\usepackage[normalem]{ulem}
\usepackage[usenames,table]{xcolor}
\usepackage[colorlinks=true,linkcolor=black,citecolor=blue,urlcolor=blue,filecolor=black]{hyperref}

\def\cC{{\cal C}}

\def\cK{{\cal K}}
\def\cL{{\cal L}}

\def\be{\begin{equation}}
\def\ee{\end{equation}}

\newcommand{\viel}{\bar{e}}

\begin{document}

\begin{frontmatter}



\title{Dual actions for massless, partially-massless and 
massive gravitons in (A)dS}

\author[UMONS]{Nicolas~Boulanger}
\author[ETH]{Andrea~Campoleoni}
\author[label4]{Ignacio~Cortese}

\address[UMONS]{Groupe de M\'ecanique et Gravitation, Unit of Theoretical and 
Mathematical Physics, Universit\'e de Mons -- UMONS, 20 place du Parc, 7000 Mons, 
Belgium\\
\href{mailto:nicolas.boulanger@umons.ac.be}{nicolas.boulanger@umons.ac.be}}
\address[ETH]{Institut f\"ur Theoretische Physik, ETH Zurich, Wolfgang-Pauli-Strasse 27, 8093 Z\"urich, Switzerland\\
\href{mailto:campoleoni@itp.phys.ethz.ch}{campoleoni@itp.phys.ethz.ch}}
\address[label4]{Departamento de F\'isica de Altas Energ\'ias, Instituto de Ciencias Nucleares -- UNAM, Circuito Exterior s/n, Cd. Universitaria, 04510 Ciudad de M\'exico, Mexico\\ \href{mailto:nachoc@nucleares.unam.mx}{nachoc@nucleares.unam.mx}}

\begin{abstract}
We provide a unified treatment of electric-magnetic duality, 
at the action level and with manifest Lorentz invariance, 
for massive, massless as well as partially-massless gravitons 
propagating in maximally symmetric spacetimes of any dimension 
$n>3\,$. 
For massive and massless fields, we complete previous analyses that use parent-action techniques by giving dual descriptions that enable direct 
counting of physical degrees of freedom in the flat and 
massless limit. The same treatment is extended to the partially-massless case, where the duality has been previously discussed in covariant form only at the level of the equations of motion. 
The nature of the dual graviton is therefore clarified for all 
values of the mass and of the cosmological constant.
\end{abstract}





\end{frontmatter}


\section{Introduction}

In this note, we complete and extend previous analyses
on dual formulations of massive and (partially) massless 
spin-2 theories in (A)dS backgrounds of arbitrary dimension 
$n>3\,$.
We resort to the parent-action technique employed in the papers
\cite{Curtright:1980yj,West:2001as,West:2002jj,Boulanger:2003vs,Matveev:2004ac,
Zinoviev:2005zj,Zinoviev:2005qp,Gonzalez:2008ar,Khoudeir:2008bu,Basile:2015jjd} 
in order to derive equivalent, dual  
actions in the sense of Fradkin and Tseytlin \cite{Fradkin:1984ai}.
In brief, in this framework one obtains two equivalent second-order 
actions ---~whose field equations are related by electric-magnetic duality~--- 
by eliminating different sets of fields from a common ``parent'' 
first-order action.
In (A)dS$_n$ these techniques have been employed for massless and massive 
gravitons, while the partially-massless case has been discussed recently 
only in $n=3$ \cite{Galviz:2017tda}. 
The same setup has also been used in the context of 
Ho{\v r}ava-Lifshitz gravity \cite{Cortese:2014pfa}.

For all values of the mass and of the cosmological constant, 
we furnish dual formulations at the action level 
and in a manifestly Lorentz-invariant way. 
The dual actions 
that we built are such that the flat and massless 
limits are smooth, thereby making the identification of the physical 
degrees of freedom and of the helicities straightforward. 
In the partially-massless case \cite{Deser:1983mm}, 
we obtain for the first 
time a dual, manifestly covariant action principle featuring 
a mixed-symmetry gauge field.

At the level of the field equations, (self-)duality symmetry, 
often named pseudo (self-)duality, 
has been studied in flat spacetime for linearised gravity
in \cite{Hull:2000zn,Hull:2001iu}; 
see also \cite{Bekaert:2002jn}.
In (A)dS$_4\,$, pseudo-duality symmetry for
partially massless spin-2 fields
was studied in \cite{Hinterbichler:2014xga,Cherney:2015jxp}. 
These are first steps towards the establishment of an 
equivalence between theories, 
for which an off-shell duality relation is necessary. 
In flat spacetime, the duality between the massless 
Fierz-Pauli action and the Curtright action 
\cite{Curtright:1980yk,Aulakh:1986cb} 
was proven in the series of works 
\cite{West:2001as,West:2002jj,Boulanger:2003vs}.

The action principles that we present feature 
both the original spin-2 field  and its dual, in a manifestly 
Lorentz-invariant fashion. On the other hand, the pair of dual fields
does not enter the action in a duality-symmetric way. 
For such a democratic appearance of electric and magnetic fields inside 
the action, the price to pay is the loss 
of manifest spacetime covariance, as explained 
for massless spin-2 and higher-spin theories around flat spacetime 
in the papers \cite{Henneaux:2004jw,Bunster:2013oaa,Henneaux:2016zlu} 
and references therein. In the same, non manifestly Lorentz-covariant 
framework, a double-potential formulation  of linearised gravity 
around (A)dS$_n$ spacetime was studied in \cite{Julia:2005ze,Leigh:2007wf} 
for $n=4$ and in \cite{Hortner:2016omi} for $n>4\,$.
As for what concerns partially-massless fields of maximal depth, 
the paper \cite{Deser:2013xb} provides an off-shell formulation 
exhibiting a nearly manifest electric-magnetic duality symmetry. 
Interestingly enough, manifest duality-invariant formulations 
of linearised gravity, in the presence of 
sources, have been given in \cite{Bunster:2006rt} and in an 
alternative way in \cite{Barnich:2008ts}; in the partially-massless
case, see also \cite{Hinterbichler:2015nua}. 
Finally, the integrability properties of duality-symmetric systems 
were studied in \cite{Barnich:2008ar}.

In more details, the unified treatment of spin-2 duality presented 
in this note leads to the following results:
\begin{itemize}
    \item In the case of a massless graviton in (A)dS, 
    we complete the programme sketched in 
    \cite{Basile:2015jjd} by linking the dual action obtained therein 
    to its Stueckelberg formulation admitting a smooth flat limit; 
    \item For partially massless spin-2 field in (A)dS, we obtain a dual 
    description at the action level, thereby elevating the duality from a pseudo to 
    a genuine off-shell duality;
    \item In the massive case in (A)dS, we clarify the flat limit of 
    the dual model presented in \cite{Zinoviev:2005zj} in that we 
    have Stueckelberg gauge fields representing the dual spin-2, 
    spin-1 and scalar sectors. 
    Therefore, our actions admit a smooth flat limit in \emph{both} 
    electric and magnetic formulations.
\end{itemize}

In sections~\ref{sec:parent} and \ref{sec:electric} 
we recall the main features 
of the first-order description of massive spin-two fields, 
that we use as a parent action. Section~\ref{sec:magnetic} 
collects our original results on dual formulations for 
spin-2 fields in (A)dS.

\section{The parent action}\label{sec:parent}

We consider as parent action the first-order Stueckelberg action describing, 
for generic values of the parameters, the propagation of a massive spin-2 
field in a constant curvature background \cite{Zinoviev:2008ze}. 
It is obtained by coupling the free actions for massless fields of spin two, 
one and zero. It thus comprises the kinetic terms for these 
fields,\footnote{We denote the background vielbein by $\viel^a\,$, 
while $\nabla$ is the Lorentz-covariant derivative on (A)dS$_n$. 
In our conventions, it satisfies $\nabla^2 V^{c} = -\,\sigma\,\lambda^2\, \viel^c \wedge \viel_b \ V^b$, so that $\sigma = 1$ in AdS$_n$ and $\sigma = -1$ in dS$_n$. 
We define the Levi-Civita symbol $\epsilon_{a_1\cdots a_n}$ such that 
$\epsilon_{01\cdots n-1}=-1$ and we adopt the mostly-plus convention for 
the metric. In the following we omit wedge products and we substitute 
groups of antisymmetrised indices with a label denoting the total number 
of indices. For instance, we introduce the 
$k$-form $H^{a[k]} \equiv H^{a_1 \cdots a_k} = \viel^{\,a_1} \cdots \viel^{\,a_k}$. 
Indices enclosed between square brackets are antisymmetrised, and dividing 
by the number of terms involved is understood (strength-one convention).
Finally, repeated indices also denote an antisymmetrisation, e.g.,
 $A_a B_a \equiv A_{[a_1} B_{a_2]}.$}
\begin{align}
\cL^{(2)} & = - \frac{\epsilon_{abcp[n-3]}}{2(n-3)!}  
\left( \omega^{ab} \nabla h^c + \frac{1}{n-2}\, \omega^a{}_q \omega^{qb} \viel^c \right) H^{p[n-3]} \, , \label{K2} \\[5pt]
\cL^{(1)} & = \frac{\epsilon_{ab c[n-2]}}{2(n-2)!}\, F^{ab} \left( \nabla A - \frac{1}{4}\, F_{kl} \, \viel^k \viel^l \right) H^{c[n-2]} \, , \label{K1} \\[5pt]
\cL^{(0)} & = \frac{\epsilon_{ab[n-1]}}{(n-1)!}\, \pi^a \left( \nabla \varphi  -\frac{1}{2} \, \pi_k \, \viel^k \right) H^{b[n-1]} \, , \label{K0}
\end{align}
together with cross couplings and mass terms:
\begin{equation}
\begin{split}
  &\cL_{\textrm{cross}} = \frac{\epsilon_{ab c[n-2]}}{(n-1)!}\, \bigg( (n-1)m\, \omega^{ab} A + m\, F^a{}_d \, h^d \viel^b + \mu\, \pi^a A\, \viel^b \\
  & -  \frac{(n-2)\mu^2}{4}\, h^a
  h^b - m\,\mu\, \varphi\, h^a \viel^b -
  \frac{m^2}{n-2}\, \varphi^2\, \viel^a \viel^b \bigg)\, H^{c[n-2]} \, .
\end{split}
\end{equation}
The full action is the integral of $\cL = \sum_{s=0}^2 \cL^{(s)} + \cL_{\textrm{cross}}$ and it is invariant under the gauge symmetries
\begin{subequations}
\begin{align}
\delta h^{a} & = \nabla \xi^a - \Lambda^{ab} \viel_b + \frac{2m}{n-2}\, \epsilon\, \viel^a \, , \label{varh} \\ 
\delta \omega^{ab} & = \nabla \Lambda^{ab} + \frac{\mu^2}{n-1}\, \viel^{[a} \xi^{b]} \, , \label{varomega}
\end{align}
\end{subequations}
and 
\begin{alignat}{5}
\delta A & = \nabla \epsilon - m\, \xi^a \viel_a  \, , \qquad
& \delta F^{ab} & = 2m\, \Lambda^{ab} \, , \label{var1} \\[5pt]
\delta \varphi & = -\,\mu\, \epsilon \, , \qquad 
& \delta \pi^a & = -\,m\,\mu\, \xi^a \, . \label{var0}
\end{alignat}
For later convenience, we introduced the constants $m$ and $\mu$, even if the action actually depends only on a single mass parameter (besides the (A)dS radius). Gauge invariance requires
\be \label{mu-m}
\mu^2 = \frac{2(n-1)}{n-2} \left(\, 2m^2 + \sigma (n-2) \lambda^2 \,\right) .
\ee

When $m=0$ the fields of spin one and zero decouple from the 
spin-two sector and one recovers the usual first-order 
formulation of linearised gravity in (A)dS. 
At $\mu = 0$, the sole scalar sector decouples and one obtains 
a first-order description of a partially-massless graviton, 
propagating helicities two and one in the flat limit. 
The first-order description for the spin-$s$ totally 
symmetric partially-massless cases of all depths 
was given in \cite{Skvortsov:2006at}.
With the manifestly unitary conventions used in \eqref{K2}--\eqref{K0}, one can set $\mu$ to zero by tuning the mass $m \in \mathbb{R}$ only in dS ($\sigma = -1$). In section \ref{sec:PM} we shall show that partially-massless fields in AdS can be described in this formalism at the price of flipping the sign of the spin-one kinetic term, which makes their lack of unitarity manifest.

Eq.~\eqref{K2} can be expressed in terms of the field \cite{West:2001as}
\be \label{defY}
Y^{bc|a} = \omega^{a|bc} + g^{ab} \omega_{d|}{}^{cd} - g^{ac} \omega_{d|}{}^{bd} \; ,
\ee
which is antisymmetric in its first two indices and transforms as
\be
\delta Y^{bc|}{}_a = \nabla_{\!a} \Lambda^{bc} + 2\, \viel_a{}^{[b} \nabla_{\!d} \Lambda^{c]d} - \frac{(n-2)\mu^2}{n-1}\,\viel_a{}^{[b}\xi^{c]} \; .
\label{varY}
\ee
The spin-2 kinetic term can then be cast in the form (from now on we will omit the integration measure $d^nx \sqrt{-g}$ brought by $\viel^{a_1} \cdots \viel^{a_n} = \det(\viel) \epsilon^{a_1 \cdots a_n} d^nx$)
\be \label{West_action}
\cL^{(2)} = \nabla_{\!b\phantom{|}\!} h_{c|}{}^a Y^{bc|}{}_a 
- \frac{1}{2} \left( Y^{bc|a} Y_{ab|c} + \frac{1}{n-2}\, Y^{ab|}{}_b Y_{ac|}{}^c \right)\, ,
\ee
while the cross couplings and mass terms read
\be \label{Lcross}
\begin{split}
&\cL_{\textrm{cross}} = -\, \frac{2m}{n-2}\, Y^{ab|}{}_b A_a - m\, F^{ab} h_{a|b} - \mu\, \pi^a A_a \\
 & - \frac{(n-2)\mu^2}{4(n-1)} \left(h_{a|b}h^{b|a} - h^2\right)  + m\,\mu\, h\,\varphi + \frac{n\,m^2}{n-2}\, \varphi^2 \, , 
\end{split}
\ee
where $h = h_{a|}{}^a$ denotes the trace of the linearised vielbein.

As recalled in section~\ref{sec:electric}, eliminating the auxiliary fields $Y^{bc|}{}_a$, $F^{ab}$ and $\pi^a$ from the parent action $\cL$ one obtains a second-order description of a massive spin-2 field in terms of the linearised metric and the fields $A_\mu$ and $\varphi$, which reduces to the Fierz-Pauli action for $m=0$. In section~\ref{sec:magnetic} we will instead show how eliminating the fields $h_{a|}{}^b$, $A_a$ and $\varphi$ leads to its dual description, involving mixed-symmetry fields for generic values of $n$. 

\section{Electric reduction}\label{sec:electric}

The equations of motion for $Y^{bc|}{}_a$, $F^{ab}$ and $\pi^a$ arising from $\cL[h,Y,A,F,\varphi,\pi]$ allow to solve for them algebraically. E.g.
\be
\begin{split}
    Y_{ab|c} &= \nabla_{\!c}h_{[a|b]}-\nabla_{\!a}h_{(b|c)}+\nabla_{\!b}h_{(a|c)}  \\
    &\phantom{==========} +2g_{c[a} \left(\nabla^d h_{b]|d}-\nabla_{\!b]} h 
    +2m\, A_{b]}\right)\, . 
\end{split}
\ee
By plugging this and the similar expressions for $F^{ab}$ and 
$\pi^a$ into the parent Lagrangian $\cL\,$, the latter reduces, 
modulo a total derivative, to the second-order Stueckelberg Lagrangian
for a symmetric spin-2 field \cite{Zinoviev:2001dt,Zinoviev:2006im}:
\be
\begin{split}\label{electricLag}
& \cL[h,A,\varphi]= -\tfrac{1}{2}\,\nabla_{\!a} h_{(b|c)}\nabla^a h^{(b|c)}  + \nabla_{\!a} h_{(b|c)}\nabla^c h^{(b|a)} \\
 & + \tfrac{1}{2}\,\nabla_{\!a} h \nabla^a h - \nabla_{\!a} h\nabla_b h^{(a|b)} -\tfrac{(n-1)\sigma\lambda^2}{2} \left(2 h_{(a|b)}h^{(a|b)}-h^2\right)  \\
 & -\nabla_{\![a}A_{b]} \nabla^{[a}A^{b]} -(n-1)\sigma\, \lambda^2 A_a A^a -\tfrac{1}{2}\, \nabla_{\!a}\varphi \nabla^a\varphi  \\ 
&-2m\, A_a\left(\nabla^a h-\nabla_{\!b} h^{(a|b)}\right) +\mu\, \varphi \nabla_{\!a} A^a \\
&- m^2\left(h_{(a|b)}h^{(a|b)}- h^2\right)  +\tfrac{n\,m^2}{n-2}\, \varphi^2 + m\,\mu\, h\,\varphi \; . 
\end{split}
\ee
The resulting action is invariant under the gauge transformations \eqref{varh} 
for $h_{(a|b)}$, to be identified with the linearised metric, together 
with \eqref{var1} and \eqref{var0} for the Stueckelberg fields $A_a$ 
and $\varphi\,$. 
The antisymmetric part of the vielbein, $h_{[a|b]}\,$, 
enters the reduced Lagrangian only through a total derivative, 
consistently with the shift symmetry it enjoys under Lorentz transformations.

The first two lines of \eqref{electricLag} gives the Fierz-Pauli Lagrangian
for a massless spin-2 field in (A)dS. 
For $\mu = 0$ one obtains a description of a partially-massless 
spin-2 field in dS in terms of the Stueckelberg coupling of the 
Fierz-Pauli and Proca Lagrangians. 
The field $A_a$ can be gauged away using $\xi^a$, and the resulting action 
is invariant under
\be
\delta h_{(a|b)} = \tfrac{1}{\lambda} \left( \nabla_{\!(a} \nabla_{\!b)} \epsilon + \lambda^2 g_{ab} \epsilon \right) \, .
\ee
In this context, 
the partially-massless gauge symmetry thus follows because  
gauge transformations \eqref{varh} and \eqref{var1} with 
\mbox{$\nabla_{\!a} \epsilon - m\, \xi_a = 0$} preserve 
the gauge fixing $A_a = 0\,$.

\section{Magnetic reduction}\label{sec:magnetic}

\subsection{Massless case}\label{sec:massless}

When $m = 0$ the fields of spin one and zero decouple and one can consider the parent Lagrangian
\be \label{Lmassless}
\begin{split}
\cL_{0}[h,Y] & =  \cL^{(2)}[h,Y] - \tfrac{(n-2)\sigma\lambda^2}{2} \left(h_{a|b}h^{b|a} - h^2\right) \, , 
\end{split}
\ee
with $\cL^{(2)}$ given in \eqref{West_action}. 
Its gauge symmetries are obtained by setting $m = 0$ in \eqref{varh} 
and \eqref{varY}. Contrary to flat space \cite{Boulanger:2003vs} 
(where it enters the action linearly), in (A)dS the linearised vielbein 
is an auxiliary field thanks to the mass term in \eqref{Lmassless}: it 
can thus be eliminated through its own equation of motion \cite{Matveev:2004ac}. 
This leads to an action depending only on the traceless projection of $Y^{bc|}{}_a\,$:
\be \label{LdualY}
\hat{Y}^{bc|}{}_a = Y^{bc|}{}_a 
+ \tfrac{2}{n-1}\,\viel_a{}^{[b} Y^{c]d|}{}_d \; . 
\ee
After the elimination of $h_{a|b}\,$, 
the trace of $Y^{bc|a}$ indeed contributes to the action only via 
a boundary term, consistently with the shift symmetry generated by $\xi^a$ 
in \eqref{varY}, which is still present for $m = 0\,$. 
One can cast the resulting Lagrangian in the form
\be 
{\cal L}_{0}[Y] = \tfrac{\sigma}{2(n-2)\lambda^2}\,\Big[
\nabla_{\!a\,} \hat{Y}^{cd|}{}_b\, \nabla_{\!c\,} \hat{Y}^{ab|}{}_d 
+ \sigma \lambda^2 \hat{Y}_{bc|a}\,\hat{Y}^{ba|c} \Big]\;,
\ee
in agreement with the result obtained by eliminating the vielbein 
from the linearised McDowell-Mansouri action \cite{Basile:2015jjd}.

Introducing the Hodge dual 
$T_{a[n-2]|b} = \tfrac12 \epsilon_{a[n-2]cd} \hat{Y}^{cd|}{}_b$ 
(which satisfies $\epsilon^{a[n-2]bc} T_{a[n-2]|b} = 0$ 
on account of $\hat{Y}^{ab|}{}_b = 0\,$), 
one obtains a dual description of a massless graviton in (A)dS$_n\,$. 
The field $T$, however, has the same structure as a massive graviton 
in flat space \cite{Curtright:1980yj}; when $\lambda = 0\,$, 
the dual of a massless spin-two field is instead a $GL(n)$ 
Young-projected\footnote{Two-column, $GL(n)$-irreducible fields 
are denoted by $[p,q]$, where $p$ and $q$ stand for the lengths of 
the first and second column of the corresponding Young tableau,  
respectively.} $[n-3,1]$ 
field \cite{West:2001as,West:2002jj,Boulanger:2003vs}. 
As discussed in \cite{Basile:2015jjd}, the different nature of the dual 
graviton in (A)dS and flat space can be explained as follows: massless 
mixed-symmetry fields display less gauge symmetries in (A)dS than in flat 
space. This is the Brink-Metsaev-Vasiliev (BMV) mechanism conjectured 
in \cite{Brink:2000ag}, proved for AdS$_n$  
in \cite{Boulanger:2008up,Boulanger:2008kw,Alkalaev:2009vm} 
and for dS$_n$ in \cite{Basile:2016aen}. It is also 
discussed in \cite{Campoleoni:2012th} from the point of view of 
reducibility conditions. 
As a result, in the flat limit, mixed-symmetry gauge fields decompose 
in multiplets of gauge fields. 
In this case, in the limit $\lambda \to 0$ the field $T$ decomposes 
into a ``proper'' $[n-3,1]$ dual graviton plus an additional field 
of type $[n-2,1]$ that does not carry any local degrees of freedom. See \cite{Basile:2015jjd,Joung:2016naf} for further comments on the role of the field $T$.

This phenomenon can be described by introducing a suitable set 
of Stueckelberg fields. In the current example, following \cite{Boulanger:2008nd}  
one can introduce a new field, antisymmetric in its first three 
indices and traceless, implementing the shift
\be \label{shiftY}
    \hat{Y}^{bc|}{}_a \, \to \, \hat{Y}^{bc|}{}_a 
    + \tfrac{1}{\lambda}\,\nabla_{\!d} W^{bcd|}{}_a\;,\quad
    W^{abc|}{}_c\equiv 0\;, 
\ee
either in the parent action \eqref{Lmassless} or in \eqref{LdualY}. 
This leads to the Lagrangian 
\be
\label{purespin2masslessandmassive}
\begin{split}
\cL_{0}[Y,W] & = 
\tfrac{1}{\lambda^2} \Big[ \tfrac{1}{2}\,\nabla_{\!c} W^{abc|d} \nabla^e W_{dbe|a} 
+ \lambda\,\hat{Y}^{ab|c} \nabla^e W_{cbe|a} \\
& + \tfrac{\sigma}{2(n-2)}\, \nabla_{\!b} 
\hat{Y}^{ab|c} \nabla^d \hat{Y}_{cd|a}  + \tfrac{\lambda^2}{2}\,
\hat{Y}^{ab|c} \hat{Y}_{ac|b} \Big] \; , \\
\end{split}
\ee
that is invariant up to total derivatives 
under\footnote{If one implements the Stueckelberg 
shift \eqref{shiftY} already in the parent action \eqref{Lmassless}, 
the vielbein acquires the new transformation 
$\delta_{\chi} h^{a|b} = \frac{n-3}{(n-2)\lambda}\, \nabla_{\!c} \chi^{abc}\,$.}
\begin{align}
\delta \hat{Y}^{bc|}{}_{a} &= \nabla_{\!d} \zeta^{bcd|}{}_{a} 
\!+\! \nabla_{\!a} \Lambda^{bc} \!+\! \tfrac{2}{n-1}\, \viel_a{}^{[b} \nabla_{\!d} \Lambda^{c]d} \!+\! \tfrac{(n-3)\lambda}{\sigma}\, \chi^{bc}{}_{a} \;,
\label{gaugevariatforYandWandhA}
\\[5pt]
\delta W^{bcd|}{}_{a} &= \nabla_{\!e} {\upsilon}^{bcde|}{}_{a}
+ \nabla_{\!a} \chi^{bcd} - \tfrac{3}{n-2}\, \viel_a{}^{[b} \nabla_{\!e} \chi^{cd]e} 
-\lambda\, \zeta^{bcd|}{}_{a} \; . \label{varW_gen}
\end{align}
Note that the new field can be gauged away using the shift symmetry 
generated by the traceless $\zeta^{bcd|a}$, while it also brings its own 
differential symmetries generated by $\upsilon^{bcde|a}$ 
(which is traceless and antisymmetric in the first four indices) 
and by the fully antisymmetric $\chi^{abc}\,$.

Introducing the Hodge dual 
$C_{a[n-3]|b} = \tfrac{1}{3!} \epsilon_{a[n-3]cde} W^{cde|}{}_b$ 
(that is a $GL(n)$ Young-projected $[n-3,1]$ field, since 
$W^{cde|}{}_b$ is traceless) and denoting 
${C'}{}_{\!a[n-4]}=C_{a[n-4]b|}{}^b$ together with
${T'}{}_{\!a[n-3]}=T_{a[n-3]b|}{}^b\,$, one obtains the dual Lagrangian
\be \label{Ldual_massless}
{\cal L}_0[C,T] = -\tfrac{1}{2\lambda^2(n-3)!} \left[ 
{\cal L}[C] +  \widehat{{\cal L}}_{\textrm{cross}} 
+  \tfrac{\sigma}{(n-2)^2}{\cal I}[T]
\right]\, ,
\ee
where (denoting antisymmetrisations with repeated indices)
\begin{align}
& {\cal L}[C] = \nabla_{\!a}C_{c[n-3]|b}\,\nabla^a C^{c[n-3]|b} 
- \nabla_{\!a} C_{b[n-3]|}{}^a \,\nabla^c C^{b[n-3]|}{}_c
\label{LagC} \\& 
- (n-3)\,\Big[ \nabla_{\!a}{C'}{}_{b[n-4]}\,\nabla^{a}{C'}^{b[n-4]} 
+\nabla_{\!b} C_{ab[n-4]|}{}^c \,\nabla^{a}C^{b[n-3]|}{}_c 
\nonumber \\ 
& -2(-1)^{n}\nabla^a C_{b[n-3]|a}\,\nabla^{b} {C'}^{b[n-4]}
- (n-4) \nabla_{\!b}{C'}{}_{\!cb[n-5]}\,\nabla^{c}{C'}^{b[n-4]} \Big]\;,
\nonumber
\end{align}
\begin{equation}
\begin{split}
    \widehat{{\cal L}}_{\textrm{cross}} = 2\lambda\,&\Big[ T_{a[n-3]b|}{}^c\,\nabla^b C^{a[n-3]|}{}_c
    - {T'}{}_{a[n-3]}\,\nabla^{b} C^{a[n-3]|}{}_b \\
    &+(-1)^n(n-3)\, {T'}{}_{\!a[n-3]}\,\nabla^{a}{C'}{}^{a[n-4]}\Big]\;,
\end{split}
\end{equation}
and
\begin{equation}
\begin{split}
    {\cal I}[T] =\;& {\cal L}[T] + \sigma(n-2)\lambda^2 \,\times
    \\
    &\,\times\, \Big[T^{a[n-2]|b}T_{a[n-2]|b} - (n-2){T'}^{a[n-3]}{T'}_{\!a[n-3]}\Big]\;.
\end{split}
\end{equation}
The expression for ${\cal L}[T]$ is obtained from ${\cal L}[C]$
in \eqref{LagC}
by replacing everywhere in the latter expression the symbols $C$ and $n$
by $T$ and $n+1\,$, respectively.

Lagrangian~\eqref{Ldual_massless} is invariant, up to total derivatives, under
\begin{align}
\delta T_{a[n-2]|b} & = (-1)^{n-1}(n-2)\, \Big[ \nabla_{\!a} \tilde{\zeta}_{a[n-3]|b} + (n-3)\sigma\lambda\, g_{ba} \tilde{\chi}_{a[n-3]} \nonumber \\
& \phantom{=} + \tfrac{(-1)^{n-1}}{n-1} \left( \nabla_{\!b} \tilde{\Lambda}_{a[n-2]} + (-1)^{n-1} \nabla_{\!a} \tilde{\Lambda}_{a[n-3]b} \right)  \Big]\;, \\[5pt]
\delta C_{a[n-3]|b} & = (-1)^{n-1}(n-3)\, \nabla_{\!a} \tilde{\upsilon}_{a[n-4]|b} 
- \lambda\, \tilde{\zeta}_{a[n-3]|b} \nonumber \\
& \phantom{=} + \tfrac{n-3}{n-2} \left( \nabla_{\!b} \tilde{\chi}_{a[n-3]} + (-1)^n \nabla_{\!a} \tilde{\chi}_{a[n-4]b} \right) \,,
\end{align}
where the parameters $\tilde{\zeta}_{a[n-3]|b}$, $\tilde{\Lambda}_{a[n-2]}\,$, 
$\tilde{\upsilon}_{a[n-4]|b}$ and $\tilde{\chi}_{a[n-3]}$ are the Hodge duals of 
those entering the transformations \eqref{gaugevariatforYandWandhA} 
and \eqref{varW_gen} (the dualisation always involves only the group of 
antisymmetrised indices).
In the limit $\lambda \to 0$ the field $T$ decouples and does not 
propagate any degrees of freedom, while one retains the gauge field
$C_{a[n-3]|b}\,$, the dual graviton in flat space~\cite{Boulanger:2003vs}.

In a spacetime of any dimension $D > n$, the action \eqref{Ldual_massless} ---~featuring one of the two possible BMV couples of fields including $T_{a[n-2]|b}$~--- would give a non-unitary propagation in dS. This is manifested by the $\sigma$-dependent relative sign between the kinetic terms that we obtained. In this specific case, the relative sign is irrelevant because $Y$ is a topological field in flat space and, indeed, the massless theory is unitary for any value of the cosmological constant. 

\subsection{Partially-massless case}\label{sec:PM}

Partially-massless spin-2 fields exist for any 
non-vanishing values of the cosmological 
constant, although they are unitary only in 
dS \cite{Higuchi:1986wu}. 
To exhibit these facts, in this subsection we slightly modify our conventions, 
multiplying $\cL^{(1)}$ by $-\sigma\,$.
With this choice the factor $\sigma$ in \eqref{mu-m} is replaced by $-1$, 
so that one can reach the point $\mu = 0$ in both dS and AdS. This leads to the parent Lagrangian
\be \label{LPM3}
\begin{split}
& \cL_{\textrm{PM}}[h,Y,A,F] = h_{a|b}\, \cC^{a|b} + \tfrac{\sigma}{\widetilde{m}}\, A_a \nabla_{\!b}\, \cC^{b|a} \\
& - \tfrac{1}{2} \left( Y^{bc|a} Y_{ab|c} + \tfrac{1}{n-2}\, Y^{ab|}{}_b Y_{ac|}{}^c  \right) - \tfrac{\sigma}{4}\, F_{ab} F^{ab} \, ,
\end{split}
\ee
where we defined
\be \label{mPM}
\cC^{a|b} = \nabla_{\!c} Y^{ac|b} - \widetilde{m}\, F^{ab} \, , \qquad
\widetilde{m} = \pm\, \lambda \sqrt{\frac{n-2}{2}} \, .
\ee
In the conventions adopted in this subsection, the gauge symmetries of the action are
\begin{align}
\delta h_{a|}{}^{b} & = \nabla_{\!a} \xi^b + \Lambda_a{}^b + \tfrac{2\widetilde{m}}{n-2}\, \viel_a{}^b \, \epsilon \, , \label{varh_PM} \\[3pt]
\delta Y^{bc|}{}_{a} &= \nabla_{\!a} \Lambda^{bc} + 2\, \viel_a{}^{[b} \nabla_{\!d} \Lambda^{c]d} \, , \\[3pt]
\delta A_a & = \nabla_{\!a} \epsilon + \sigma\, \widetilde{m}\, \xi_a \, , \\[3pt]
\delta F^{ab} & = -\,2\,\sigma\,\widetilde{m}\, \Lambda^{ab} \, . \label{varF_PM}
\end{align}

In \eqref{LPM3} we stressed that the fields $h_{a|b}$ and $A_a$ are both Lagrange multipliers when $\mu = 0$ (although the constraint imposed by the latter field is not independent). The analysis of the partially-massless case therefore follows closely that of a massless graviton in flat space \cite{Boulanger:2003vs}, rather than those presented in sections \ref{sec:massless} and \ref{sec:massive}. The constraint \mbox{$\cC_{a|b} = 0$} is solved by
\be \label{solY}
Y^{bc|}{}_a = \frac{1}{\lambda}\, \nabla_{\!d} W^{bcd|}{}_a - \frac{\sigma}{2\widetilde{m}} \left( \nabla_{\!a} F^{bc} + 2\, \viel_a{}^{[b} \nabla_{\!d} F^{c]d} \right) \,,
\ee
where $W^{bcd|}{}_a$ has the same structure as the field introduced in the 
Stueckelberg shift~\eqref{shiftY}. In particular, it is traceless.
Substituting \eqref{solY} in \eqref{LPM3}, one obtains
\be \label{LPM_fin}
\begin{split}
& \cL_{\textrm{PM}}[W] = -\,\frac{1}{2\lambda^2}\,\nabla_{\!d} W^{bcd|a} \nabla^{e} W_{abe|c} \\
& + \nabla_{\!a} \left( F^{ab} \nabla^{c} F_{bc} - F_{bc} \nabla^{c} F^{ab} + \tfrac{4\sigma\widetilde{m}}{\lambda}\, F_{bc} \nabla_{\!d} W^{abd|c} \right) \,.
\end{split}
\ee
This Lagrangian actually depends only on the field 
$W^{bcd|a}$: $F^{ab}$ contributes only via a total derivative consistently with the shift symmetry \eqref{varF_PM}. It is still invariant under
\be
\delta W^{bcd|}{}_{a} = \nabla_{\!e} {\upsilon}^{bcde|}{}_{a} \, ,
\ee
while the other differential symmetry that was present in the massless case (cf.~\eqref{varW_gen}) is absent.

All gauge symmetries that the field $W^{bcd|a}$ and, consequently, 
its Hodge dual would display in flat space can be recovered by implementing the 
Stueckelberg shift
\be
W^{bcd|}{}_a \to W^{bcd|}{}_a + \frac{\widetilde{m}^{-1}}{n-3}\, 
\left( \nabla_{\!a} U^{bcd} - \tfrac{3}{n-2}\, \viel_a{}^{[b} 
\nabla_{\!e} U^{cd]e} \right) \,.
\ee
Substituting in \eqref{LPM_fin} one obtains the Lagrangian
\be
\begin{split}
& \cL_{\textrm{PM}}[W,U] =  -\,\tfrac{1}{2\lambda^2}\,\nabla_{\!d} W^{bcd|a} \nabla^{e} 
W_{abe|c} + \tfrac{\sigma}{\widetilde{m}} \, U_{abc} \nabla_{\!d} W^{abd|c} \\
& - \tfrac{\sigma}{2(n-2)\widetilde{m}^2}\, \nabla_{\!c} U^{abc} \nabla^d U_{abd} - \tfrac{\lambda^2}{2\widetilde{m}^2}\, U_{abc} U^{abc} \, ,
\end{split}
\ee
which is invariant up to total derivatives under
\begin{align}
\delta W^{bcd|}{}_a & = \nabla_{\!e} \upsilon^{bcde|}{}_a \!+\! \nabla_{\!a} \chi^{bcd} \!-\! \tfrac{3}{n-2}\, \viel_a{}^{[b} \nabla_{\!e} \chi^{cd]e} \!-\! \tfrac{\sigma\lambda^2}{\widetilde{m}}\, \rho^{bcd}{}_a \, , \label{varPMW} \\[5pt]
\delta U^{abc} & = \nabla_{\!d} \rho^{abcd} - (n-3)\,\widetilde{m}\, \chi^{abc} \, . \label{varPMU}
\end{align}
The contribution in $\rho$ in \eqref{varPMW} (that was absent in \eqref{varW_gen}) is necessary because, contrary to the massless case, the field $U$ does not enter the action only via its divergence.

As in the massless case, the sign of one of the two kinetic terms depends on $\sigma$. 
This is consistent with the observation that, after Hodge dualisation, 
one obtains a BMV couple of fields which is unitary only in 
dS \cite{Basile:2016aen}. 
However, in this case both fields propagate in the flat limit: 
the $[n-3,1]$ dual of $W$ carries the spin-2 helicities, 
while the $[n-3]$ dual of $U$ carries the spin-1 helicities. 
Consequently, the sign flip of a kinetic terms does matter: recovering 
the BMV couple of fields that is not-unitary in AdS is just another 
way to see that partially-massless fields are not unitary in AdS.

Using the dual field $C_{a[n-3]|b}$ defined as in the massless case
and introducing the Hodge dual field
${A}_{a[n-3]} = \tfrac{1}{3!} \epsilon_{a[n-3]bcd} U^{bcd}\,$, 
the Stueckelberg Lagrangian we obtain for the dual partially massless 
spin-2 field in (A)dS$_n$ is
\begin{align}
    {\cal L}_{\textrm{PM}} =  -\tfrac{1}{2(n-3)!\lambda^2}\Big[ 
{\cal L}[C]  \,-\, & \tfrac{2\sigma \lambda^2}{(n-2)\widetilde{m}^2}\, 
{\cal L}[A] 
+ \tfrac{4\sigma\lambda^2}{\widetilde{m}} 
\widetilde{\cal L}_{\textrm{cross}} \Big]\;,
\end{align}
where ${\cal L}[C]$ is given in \eqref{LagC}, 
\begin{align}
    {\cal L}[A] = \nabla_{\!a} A_{b[n-3]}\,\nabla^a A^{b[n-3]}
    \,-\,&(n-3)\nabla^a{A}_{c[n-4]a}\,\nabla_{\!b}{A}^{c[n-4]b}
    \nonumber \\
    & + 3\sigma \lambda^2 \,{A}_{a[n-3]} {A}^{a[n-3]}\;,
\end{align}
and the cross terms are
\begin{equation}
    \widetilde{\cal L}_{\textrm{cross}} =  
    {A}^{a[n-3]} \left( \nabla_{\!b}C_{a[n-3]|}{}^b
    + (-1)^{n-1} (n-3)\,\nabla_{\!a}{C'}{}_{a[n-2]} \right) .
\end{equation}
The action is invariant under
\begin{align}
\delta C_{a[n-3]|b} & = (-1)^{n-1}(n-3) \left( \nabla_{\!a} \tilde{\upsilon}_{a[n-4]|b} - \tfrac{\sigma\lambda^2}{\widetilde{m}}\, g_{ba}\, \tilde{\rho}_{a[n-4]} \right) \nonumber \\
& \phantom{=} + \tfrac{n-3}{n-2} \left( \nabla_{\!b} \tilde{\chi}_{a[n-3]} + (-1)^n \nabla_{\!a} \tilde{\chi}_{a[n-4]b} \right) , \\[5pt]
\delta A_{a[n-3]} & = (n-3) \left( (-1)^{n-1}\nabla_{\!a} \tilde{\rho}_{a[n-4]} 
- \widetilde{m}\, \tilde{\chi}_{a[n-3]} \right) ,
\end{align}
where the parameters $\tilde{\upsilon}_{a[n-4]|b}$, $\tilde{\chi}_{a[n-3]}$ and $\tilde{\rho}_{a[n-4]}$ are the Hodge duals of those entering the transformations \eqref{varPMW} and \eqref{varPMU}.

\subsection{Massive case}\label{sec:massive}

We now consider the full Stueckelberg action presented in section \ref{sec:parent}. The elimination of the fields $h_{a|b}$, $A_a$ and $\varphi$ has been considered in \cite{Zinoviev:2005zj,Khoudeir:2008bu}. In the spirit of our discussion of the special points $m = 0$ and $\mu = 0$, we complement these works by exhibiting a dual description with a smooth massless and flat limit. For generic values of $m$, $h_{a|b}$ is an auxiliary field and it can be eliminated through its equation of motion as in section~\ref{sec:massless}. The field $A_a$ is instead a Lagrange multiplier enforcing the constraint
\be
\nabla_{\!b} F^{ba} - \tfrac{2m}{n-2}\, Y^{ab|}{}_b - \mu\,\pi^a = 0 \, ,
\ee
which can be solved by expressing $\pi^a$ in terms of the other fields. 
The equation of motion of $\varphi$ does not bring any new information, 
since it is not independent on account of the Noether identity associated 
with the gauge symmetry generated by $\epsilon\,$.

Substituting the on-shell values of $h_{a|b}$ and $\pi^a$ 
in the Stueckelberg Lagrangian leads to \cite{Zinoviev:2005zj}
\begin{align}
& \cL[\hat{Y},F] = \tfrac{1}{\mu^2} \left[ \tfrac{n-1}{n-2}\, \nabla_{\!b} \hat{Y}^{ab|c} \nabla^d \hat{Y}_{cd|a} + \tfrac{\mu^2}{2}\, \hat{Y}^{ab|c} \hat{Y}_{ac|b} \right.  \label{Lag2massive} \\ 
& \left. + \tfrac{1}{2}\, \nabla_{\!b} F^{ab} \nabla^c F_{ac} - \tfrac{2(n-1)m}{n-2}\, F_{ab} \nabla_{\!c} \hat{Y}^{bc|a} + \left( \tfrac{\mu^2}{4} - \tfrac{(n-1)m^2}{n-2} \right) F_{ab} F^{ab} \right] , \nonumber
\end{align}
where we recall that the parameters $m$ and $\mu$ are related by \eqref{mu-m}. 
One can then consider the Hodge duals of the fields $\hat{Y}$ and $F$ 
and obtain a dual theory for a massive graviton in terms of the Stueckelberg 
coupling of a massless spin-2 field (accounted by the $[n-2,1]$ dual of $\hat{Y}$) 
with a Proca field (accounted by the $[n-2]$ dual of $F$). 
Its gauge symmetries are those inherited from \eqref{var1} and 
\eqref{varY} after dualisation.

In order to obtain a smooth massless and flat limit, one should introduce 
two additional fields: the traceless $W^{bcd|}{}_a$ that we already encountered 
in section \ref{sec:massless} and a 3-form $U^{abc}\,$. 
This will allow to recover all Curtright gauge symmetries for the Hodge dual 
of $\hat{Y}^{bc|}{}_a$ and the usual gauge symmetry for the massless $(n-2)\,$-form 
which is the Hodge dual of $F^{ab}\,$. 
Due to the coupling $F^{ab} h_{a|b}$ in \eqref{Lcross}, 
introducing the 3-form via a Stueckelberg shift of $F^{ab}$ would modify 
the equation of motion for $h_{a|b}$ and, as a result, it would introduce 
second-order kinetic terms mixing $U^{abc}$ with the fully antisymmetric 
projection of $Y^{ab|c}\,$. On the other hand, the shifts
\begin{subequations} \label{shift_massive}
\begin{align} 
Y^{bc|a} & \to Y^{bc|a} + \tfrac{1}{\mu}\, \nabla_{\!d} W^{bcd|a} - \tfrac{m}{\mu}\, U^{abc} \, , \\
F^{ab} & \to F^{ab} + \tfrac{1}{\mu}\, \nabla_{\!c} U^{abc}
\end{align}
\end{subequations}
do not modify the equation of motion for $h_{a|b}$ and therefore they cannot introduce any mixed kinetic term. The elimination of the fields $h_{a|b}$, $A_a$ and $\varphi$ then proceeds as above and one obtains the sum of the kinetic terms
\be \label{kinetic_massive}
\begin{split}
\cK =\; & \tfrac{1}{\mu^2} 
\Big[
\tfrac{1}{2}\,\nabla_{\!c} W^{abc|d} \nabla^e W_{dbe|a}
 + \tfrac{n-1}{n-2}\, \nabla_{\!b} \hat{Y}^{ab|c} \nabla^d \hat{Y}_{cd|a} \\
& + \tfrac{1}{4}\, \nabla_{\!c} U^{abc} \nabla^d U_{abd} + \tfrac{1}{2}\, 
\nabla_{\!b} F^{ab} \nabla^c F_{ac} \Big]\;,
\end{split}
\ee
with the cross couplings
\be \label{Lcross_massive}
\begin{split}
 \cL^{(1)}_{\textrm{cross}} =\; & \tfrac{1}{\mu}\, \Big[ \hat{Y}_{ab|c} 
 \nabla_{\!d} W^{acd|b} + \tfrac{m}{\mu}\, U_{abc} \nabla_{\!d} W^{abd|c} \\
& - \tfrac{2(n-1)m}{(n-2)\mu}\, F_{ab} \nabla_{\!c} \hat{Y}^{bc|a} 
+ \tfrac{1}{2}\, F_{ab} \nabla_{\!c} U^{abc} \Big]
\end{split}
\ee
and mass-like terms
\be \label{Lmass_massive}
\begin{split}
& \cL^{(2)}_{\textrm{cross}} = \tfrac{1}{2}\, 
\hat{Y}^{ab|c} \hat{Y}_{ac|b} + \tfrac{m}{\mu}\, \hat{Y}^{ab|c} U_{abc} \\
& - \tfrac{m^2}{2\mu^2}\, U^{abc} U_{abc} 
+ \left( \tfrac{1}{4} - \tfrac{(n-1)m^2}{(n-2)\mu^2} \right) 
F^{ab} F_{ab} \;.
\end{split}
\ee
This Lagrangian is invariant up to total derivatives under the following gauge transformations:\footnote{Implementing the shift \eqref{shift_massive} before the elimination of $h_{a|b}$ etc.\ from the parent action does not modify the gauge transformations of $A_a$, $\varphi$ and $\pi^a$. The variation of $h_{a|b}$ takes instead the same form as in the massless case and acquires a contribution $\delta_{\chi} h^{a|b} = \frac{n-3}{(n-2)\mu}\, \nabla_{\!c} \chi^{abc}$.}
\begin{align}
\delta W^{bcd|}{}_a & = \nabla_{\!e} {\upsilon}^{bcde|}{}_{a}
+ \nabla_{\!a} \chi^{bcd} - \tfrac{3}{n-2}\, \viel_a{}^{[b} \nabla_{\!e} \chi^{cd]e} \nonumber \\
& \phantom{=} -\mu\, \zeta^{bcd|}{}_{a} - m\, \rho^{bcd}{}_{a} \, , \label{varWmassive}\\[5pt]
\delta \hat{Y}^{bc|}{}_a & = \nabla_{\!d} \zeta^{bcd|}{}_{a} + \nabla_{\!a} \Lambda^{bc} + \tfrac{2}{n-1}\, \viel_a{}^{[b} \nabla_{\!d} \Lambda^{c]d} \nonumber \\
& \phantom{=} + \tfrac{(n-3)\mu}{2(n-1)}\,\chi^{bc}{}_{a} - m\, \psi^{bc}{}_a \, , \label{varYmassive} \\[5pt]
\delta U^{abc} & = \nabla_{\!d} \rho^{abcd} - \mu\, \psi^{abc} + \tfrac{2(n-3)m}{n-2}\, \chi^{abc}  \, , \label{varUmassive} \\[5pt]
\delta F^{ab} & = \nabla_{\!c} \psi^{abc} + 2m\, \Lambda^{ab} \, . \label{varFmassive}
\end{align}

The action involving the Hodge duals of the previous fields now admits a 
smooth flat and massless limit, in which different helicities decouple. 
The spin-two ones are carried by the $[n-3,1]$ Hodge dual of $W$ (as discussed 
in section~\ref{sec:massless}), while spin-one and zero helicities are carried, 
respectively, by the fully-antisymmetric Hodge duals of $U$ and $F\,$. 
We refrain from displaying this action explicitly, as it can  
straightforwardly be obtained by expressing all fields in terms of their Hodge 
duals in \eqref{kinetic_massive}--\eqref{Lmass_massive}. One can also check 
that the appropriate Curtright gauge symmetries are recovered from 
\eqref{varWmassive}--\eqref{varFmassive} together with their gauge-for-gauge 
symmetries.

\section*{Acknowledgments}
We are grateful to Th.~Basile for discussions and collaboration at an early 
stage of this work.
We thank X.~Bekaert, J.~A.~Garc\'ia and L.~Traina for fruitful discussions. 
We performed or checked several computations with the package xTras \cite{Nutma:2013zea} 
of the suite of Mathematica packages xAct. 
N.B.\ thanks ETH Z\"urich and the Institut Denis Poisson (Universit\'e de Tours),
while A.C.\ thanks the Universit\`a di Firenze and INFN (sezione di Firenze),  
for kind hospitality during the completion of this paper. 
The stay of N.B.\ at the Institut Denis Poisson in Tours was
funded by a grant of the Universit\'e de Mons (UMONS). 
The work of N.B.\ has been supported in part by a FNRS PDR grant (number T.1025.14), while the work of A.C.\ has been supported in part by the NCCR SwissMAP, funded by the Swiss National Science Foundation.



\begin{thebibliography}{10}
\expandafter\ifx\csname url\endcsname\relax
  \def\url#1{\texttt{#1}}\fi
\expandafter\ifx\csname urlprefix\endcsname\relax\def\urlprefix{URL }\fi
\expandafter\ifx\csname href\endcsname\relax
  \def\href#1#2{#2} \def\path#1{#1}\fi

\bibitem{Curtright:1980yj}
T.~L. Curtright, P.~G.~O. Freund, {Massive Dual Fields}, Nucl. Phys.
\newblock  B172 (1980) 413--424.

\bibitem{West:2001as}
P.~C. West, {E(11) and M theory}, Class. Quant. Grav. 18 (2001) 4443--4460.
\newblock \href {http://arxiv.org/abs/hep-th/0104081}
  {\path{arXiv:hep-th/0104081}}.

\bibitem{West:2002jj}
P.~C. West, {Very extended E(8) and A(8) at low levels, gravity and
  supergravity}, Class. Quant. Grav. 20 (2003) 2393--2406.
\newblock \href {http://arxiv.org/abs/hep-th/0212291}
  {\path{arXiv:hep-th/0212291}}.

\bibitem{Boulanger:2003vs}
N.~Boulanger, S.~Cnockaert, M.~Henneaux, {A note on spin s duality}, JHEP 06
  (2003) 060.
\newblock \href {http://arxiv.org/abs/hep-th/0306023}
  {\path{arXiv:hep-th/0306023}}.

\bibitem{Matveev:2004ac}
A.~S. Matveev, M.~A. Vasiliev, {Dual formulation for higher spin gauge
  fields in (A)dS$_d$}, Phys. Lett. B609 (2005) 157--166.
\newblock \href {http://arxiv.org/abs/hep-th/0410249}
  {\path{arXiv:hep-th/0410249}}.

\bibitem{Zinoviev:2005zj}
{\relax Yu}.~M. Zinoviev, {On dual formulations of massive tensor fields}, JHEP
  10 (2005) 075.
\newblock \href {http://arxiv.org/abs/hep-th/0504081}
  {\path{arXiv:hep-th/0504081}}.

\bibitem{Zinoviev:2005qp}
{\relax Yu}.~M. Zinoviev, {On dual formulation of gravity}, \href
  {http://arxiv.org/abs/hep-th/0504210} {\path{arXiv:hep-th/0504210}}.

\bibitem{Gonzalez:2008ar}
B.~Gonzalez, A.~Khoudeir, R.~Montemayor, L.~F. Urrutia, {Duality for massive
  spin two theories in arbitrary dimensions}, JHEP 09 (2008) 058.
\newblock \href {http://arxiv.org/abs/0806.3200} {\path{arXiv:0806.3200}}.

\bibitem{Khoudeir:2008bu}
A.~Khoudeir, R.~Montemayor, L.~F. Urrutia, {Dimensional reduction as a method
  to obtain dual theories for massive spin two in arbitray dimensions}, Phys.
  Rev. D78 (2008) 065041.
\newblock \href {http://arxiv.org/abs/0806.4558} {\path{arXiv:0806.4558}}.

\bibitem{Basile:2015jjd}
T.~Basile, X.~Bekaert, N.~Boulanger, {Note about a pure spin-connection
  formulation of general relativity and spin-2 duality in (A)dS}, Phys. Rev.
  D93~(12) (2016) 124047.
\newblock \href {http://arxiv.org/abs/1512.09060} {\path{arXiv:1512.09060}}.

\bibitem{Fradkin:1984ai}
E.~S. Fradkin, A.~A. Tseytlin, {Quantum Equivalence of Dual Field Theories},
  Annals Phys.
\newblock  162 (1985) 31.

\bibitem{Galviz:2017tda}
D.~Galviz, A.~Khoudeir, {Partially Massless and Self Duality in three
  dimensions}, \href {http://arxiv.org/abs/1712.01050} {\path{arXiv:1712.01050}}.

\bibitem{Cortese:2014pfa}
I.~Cortese, J.~A. Garc{\'i}a, {Electric-magnetic duality in linearized Ho{\v
  r}ava-Lifshitz gravity}, Phys. Rev. D90~(6) (2014) 064029.
\newblock \href {http://arxiv.org/abs/1405.6536} {\path{arXiv:1405.6536}}.

\bibitem{Deser:1983mm}
S.~Deser, R.~I. Nepomechie, {Gauge Invariance Versus Masslessness in De Sitter
  Space}, Annals Phys.
\newblock  154 (1984) 396.

\bibitem{Hull:2000zn}
C.~M. Hull, {Strongly coupled gravity and duality}, Nucl. Phys. B583 (2000)
  237--259.
\newblock \href {http://arxiv.org/abs/hep-th/0004195}
  {\path{arXiv:hep-th/0004195}}.

\bibitem{Hull:2001iu}
C.~M. Hull, {Duality in gravity and higher spin gauge fields}, JHEP 09 (2001)
  027.
\newblock \href {http://arxiv.org/abs/hep-th/0107149}
  {\path{arXiv:hep-th/0107149}}.

\bibitem{Bekaert:2002jn}
X.~Bekaert, N.~Boulanger, {Massless spin two field S duality}, Class. Quant.
  Grav. 20 (2003) S417--S424.
\newblock \href {http://arxiv.org/abs/hep-th/0212131}
  {\path{arXiv:hep-th/0212131}}.

\bibitem{Hinterbichler:2014xga}
K.~Hinterbichler, {Manifest Duality Invariance for the Partially Massless
  Graviton}, Phys. Rev. D91~(2) (2015) 026008.
\newblock \href {http://arxiv.org/abs/1409.3565} {\path{arXiv:1409.3565}}.

\bibitem{Cherney:2015jxp}
D.~Cherney, S.~Deser, A.~Waldron, G.~Zahariade, {Non-linear duality invariant
  partially massless models?}, Phys. Lett. B753 (2016) 293--296.
\newblock \href {http://arxiv.org/abs/1511.01053} {\path{arXiv:1511.01053}}.

\bibitem{Curtright:1980yk}
T.~Curtright, {Generalized gauge fields}, Phys. Lett.
\newblock  165B (1985) 304--308.

\bibitem{Aulakh:1986cb}
C.~S. Aulakh, I.~G. Koh, S.~Ouvry, {Higher Spin Fields With Mixed Symmetry},
  Phys. Lett.
\newblock  B173 (1986) 284--288.

\bibitem{Henneaux:2004jw}
M.~Henneaux, C.~Teitelboim, {Duality in linearized gravity}, Phys. Rev. D71
  (2005) 024018.
\newblock \href {http://arxiv.org/abs/gr-qc/0408101}
  {\path{arXiv:gr-qc/0408101}}.

\bibitem{Bunster:2013oaa}
C.~Bunster, M.~Henneaux, S.~H{\"o}rtner, {Twisted Self-Duality for Linearized
  Gravity in D dimensions}, Phys. Rev. D88~(6) (2013) 064032.
\newblock \href {http://arxiv.org/abs/1306.1092} {\path{arXiv:1306.1092}}.

\bibitem{Henneaux:2016zlu}
M.~Henneaux, S.~H{\"o}rtner, A.~Leonard, {Twisted self-duality for higher spin
  gauge fields and prepotentials}, Phys. Rev. D94~(10) (2016) 105027, [Erratum:
  Phys. Rev.D97,no.4,049901(2018)].
\newblock \href {http://arxiv.org/abs/1609.04461} {\path{arXiv:1609.04461}}.

\bibitem{Julia:2005ze}
B.~Julia, J.~Levie, S.~Ray, {Gravitational duality near de Sitter space}, JHEP
  11 (2005) 025.
\newblock \href {http://arxiv.org/abs/hep-th/0507262}
  {\path{arXiv:hep-th/0507262}}.
  
\bibitem{Leigh:2007wf}
R.~G.~Leigh, A.~C.~Petkou,
  {Gravitational duality transformations on (A)dS(4)},
  JHEP 11 (2007) 079.
\newblock \href {http://arxiv.org/abs/0704.0531} {\path{arXiv:0704.0531}}.

\bibitem{Hortner:2016omi}
S.~H{\"o}rtner, {A deformation of the Curtright action}, Phys. Rev. D95~(2)
  (2017) 024039.
\newblock \href {http://arxiv.org/abs/1610.07225} {\path{arXiv:1610.07225}}.

\bibitem{Deser:2013xb}
S.~Deser, A.~Waldron, {PM = EM: Partially Massless Duality Invariance}, Phys.
  Rev. D87 (2013) 087702.
\newblock \href {http://arxiv.org/abs/1301.2238} {\path{arXiv:1301.2238}}.

\bibitem{Bunster:2006rt}
C.~W. Bunster, S.~Cnockaert, M.~Henneaux, R.~Portugues, {Monopoles for
  gravitation and for higher spin fields}, Phys. Rev. D73 (2006) 105014.
\newblock \href {http://arxiv.org/abs/hep-th/0601222}
  {\path{arXiv:hep-th/0601222}}.

\bibitem{Barnich:2008ts}
G.~Barnich, C.~Troessaert, {Manifest spin 2 duality with electric and magnetic
  sources}, JHEP 01 (2009) 030.
\newblock \href {http://arxiv.org/abs/0812.0552} {\path{arXiv:0812.0552}}.

\bibitem{Hinterbichler:2015nua}
K.~Hinterbichler, R.~A. Rosen, {Partially Massless Monopoles and Charges},
  Phys. Rev. D92~(10) (2015) 105019.
\newblock \href {http://arxiv.org/abs/1507.00355} {\path{arXiv:1507.00355}}.

\bibitem{Barnich:2008ar}
G.~Barnich, C.~Troessaert, {Duality and integrability: Electromagnetism,
  linearized gravity and massless higher spin gauge fields as bi-Hamiltonian
  systems}, J. Math. Phys. 50 (2009) 042301.
\newblock \href {http://arxiv.org/abs/0812.4668} {\path{arXiv:0812.4668}}.

\bibitem{Zinoviev:2008ze}
{\relax Yu}.~M. Zinoviev, {Frame-like gauge invariant formulation for massive
  high spin particles}, Nucl. Phys. B808 (2009) 185--204.
\newblock \href {http://arxiv.org/abs/0808.1778} {\path{arXiv:0808.1778}}.

\bibitem{Skvortsov:2006at}
  E.~D.~Skvortsov, M.~A.~Vasiliev,
  {Geometric formulation for partially massless fields},
  Nucl.\ Phys.\ B {\bf 756} (2006) 117.
  \newblock \href {http://arxiv.org/abs/hep-th/0601095}
 {\path{arXiv:hep-th/0601095}}.

\bibitem{Zinoviev:2001dt}
{\relax Yu}.~M. Zinoviev, {On massive high spin particles in AdS}, \href
  {http://arxiv.org/abs/hep-th/0108192} {\path{arXiv:hep-th/0108192}}.

\bibitem{Zinoviev:2006im}
{\relax Yu}.~M. Zinoviev, {On massive spin 2 interactions}, Nucl. Phys. B770
  (2007) 83--106.
\newblock \href {http://arxiv.org/abs/hep-th/0609170}
  {\path{arXiv:hep-th/0609170}}.

\bibitem{Brink:2000ag}
L.~Brink, R.~R. Metsaev, M.~A. Vasiliev, {How massless are massless fields in
  AdS(d)}, Nucl. Phys. B586 (2000) 183--205.
\newblock \href {http://arxiv.org/abs/hep-th/0005136}
  {\path{arXiv:hep-th/0005136}}.

\bibitem{Boulanger:2008up}
N.~Boulanger, C.~Iazeolla, P.~Sundell, {Unfolding Mixed-Symmetry Fields in AdS
  and the BMV Conjecture: I. General Formalism}, JHEP 07 (2009) 013.
\newblock \href {http://arxiv.org/abs/0812.3615} {\path{arXiv:0812.3615}}.

\bibitem{Boulanger:2008kw}
N.~Boulanger, C.~Iazeolla, P.~Sundell, {Unfolding Mixed-Symmetry Fields in AdS
  and the BMV Conjecture. II. Oscillator Realization}, JHEP 07 (2009) 014.
\newblock \href {http://arxiv.org/abs/0812.4438} {\path{arXiv:0812.4438}}.

\bibitem{Alkalaev:2009vm}
K.~B. Alkalaev, M.~Grigoriev, {Unified BRST description of AdS gauge fields},
  Nucl. Phys. B835 (2010) 197--220.
\newblock \href {http://arxiv.org/abs/0910.2690} {\path{arXiv:0910.2690}}.

\bibitem{Basile:2016aen}
T.~Basile, X.~Bekaert, N.~Boulanger, {Mixed-symmetry fields in de Sitter space:
  a group theoretical glance}, JHEP 05 (2017) 081.
\newblock \href {http://arxiv.org/abs/1612.08166} {\path{arXiv:1612.08166}}.

\bibitem{Campoleoni:2012th}
A.~Campoleoni, D.~Francia, {Maxwell-like Lagrangians for higher spins}, JHEP 03
  (2013) 168.
\newblock \href {http://arxiv.org/abs/1206.5877} {\path{arXiv:1206.5877}}.

\bibitem{Joung:2016naf}
E.~Joung and K.~Mkrtchyan, {Weyl Action of Two-Column Mixed-Symmetry Field and Its Factorization Around (A)dS Space}, JHEP 06 (2016) 135.
\newblock \href {http://arxiv.org/abs/1604.05330} {\path{arXiv:1604.05330}}.

\bibitem{Boulanger:2008nd}
N.~Boulanger, O.~Hohm, {Non-linear parent action and dual gravity}, Phys. Rev.
  D78 (2008) 064027.
\newblock \href {http://arxiv.org/abs/0806.2775} {\path{arXiv:0806.2775}}.

\bibitem{Higuchi:1986wu}
A.~Higuchi, {Symmetric Tensor Spherical Harmonics on the $N$ Sphere and Their
  Application to the De Sitter Group SO($N$,1)}, J. Math. Phys. 28 (1987) 1553,
\newblock [Erratum: J. Math. Phys.43,6385(2002)].

\bibitem{Nutma:2013zea}
T.~Nutma, {xTras : A field-theory inspired xAct package for mathematica},
  Comput. Phys. Commun. 185 (2014) 1719--1738.
\newblock \href {http://arxiv.org/abs/1308.3493} {\path{arXiv:1308.3493}}.

\end{thebibliography}





\end{document}